\documentclass[final,numbers,3p,12pt,times]{elsarticle}

\usepackage{amssymb,bm,amsmath}

\usepackage[nodots]{numcompress}

\biboptions{sort&compress}
\biboptions{sort&compress,comma,authoryear,numbers}

\begin{document}

\begin{frontmatter}

\title{Large Eddy Simulation of urban boundary layer flows using a canopy stress method} 
\author[ref1]{Jahrul M Alam\corref{cor1}}
\ead{alamj@mun.ca}
\cortext[cor1]{Corresponding author}
\author[ref1]{Luke P. J. Fitzpatrick}

\address[ref1]{Department of Mathematics and Statistics, Memorial University, Canada, A1C 5S7}

\begin{abstract}
  Large-eddy simulation (LES) of a turbulent flow through an array of building-like obstacles is an idealized model to study transport of contaminants in the urban atmospheric boundary layer (UABL). A reasonably accurate LES prediction of turbulence in such an UABL must resolve a significant proportion of the small but energetic eddies in the roughness sublayer, which remains prohibitive even though computational power has been increased significantly. Recently, some researchers reported a high level of inaccuracy in turbulence prediction if LES were coupled with an adaptive mesh refinement technique to resolve the roughness sublayer with a near optimal computational cost. In this article, we present a large-eddy simulation methodology to study turbulence in UABLs, where the turbulence closure is based on coupling the eddy viscosity method with the canopy stress method. Unlike  the classical Smagorinsky model that considers only the `strain portion' of the velocity gradient tensor, we consider both the `strain tensor' and the `rotation tensor' to compute the eddy viscosity. This allows us to dynamically adapt the rate of energy dissipation to the scales of the energetic eddies in the roughness sublayer. Without employing a mesh conforming to the urban roughness elements, the effect of such solid bodies are represented in the LES model through a canopy stress method in which the loss of pressure and the sink of momentum due to the interaction between eddies and roughness elements are parameterized using the instantaneous velocity field. Simulation results of the proposed canopy stress method is compared with that of a conventional Computational Fluid Dynamics (CFD) method employing a block-structured mesh conforming around the roughness elements. For urban flow simulations, the results demonstrate that the proposed canopy stress model is accurate in predicting vertical profiles of mean and variance, as well as the temporal intermittency of coherent structures. 
\end{abstract}


\end{frontmatter}

\section{Introduction}
\begin{figure}[b]
  \centering
  \includegraphics[height=12cm]{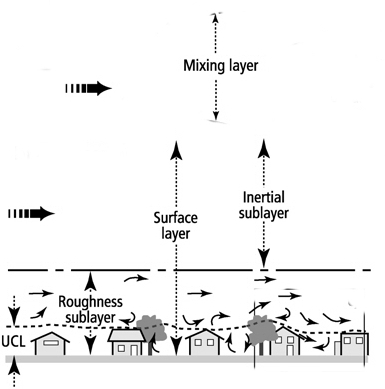}
  \caption{A schematic description of the vertical structure of urban atmospheric urban boundary layer. The depth of the viscous sublayer is typically less than 5 wall units, which is about $0.25$~mm with respect to a friction velocity of $40$~cm/s. The depth of the urban canopy layer is the mean building height, $H$. The roughness sublayer extends a depth of about $2H$ to $5H$. The surface layer usually extends up to the bottom $10$\% of the ABL, where the inertial sublayer is the part of the surface layer that overlies the roughness sublayer.}
  \label{fig:uabl}
\end{figure}

Land-surface modification in the urban atmospheric boundary~(UABL) layer have attracted numerous research efforts and practical concerns for its direct and indirect impacts on many processes behind global changes, such as increased Greenhouse Gas (GHG) concentrations, increased water and energy demand, environmental pollution etc~\cite[][]{Grimmond2011}. There is currently a growing trend of land-surface modification and urbanization because over half the world's population lives in urban areas, which is projected to be two third by $2045$~\cite[][]{Resler2017}. Since urban dwellers are exposed to an increased level of air pollution with detrimental effects on health and comfort, mixing and transport of pollutants by turbulence within urban areas remain a critical research topic~\cite[][]{Castro2009,Grimmond2011,Resler2017}. As it is illustrated schematically in Fig~\ref{fig:uabl}, the bottom $10$\% of the UABL is the surface layer above which the outer layer takes the form of either the mixed layer or the Ekman layer. Mesoscale effects of pressure gradient force, Coriolis force, Earth's rotation, and convection processes become important to govern turbulence in the outer layer. For a neutral  stratification with Coriolis parameter $f=10^{-4}\hbox{s}^{-1}$ and surface friction velocity $u_*= 40$~cm/s, the top of the outer layer may be at a distance of $1.5$~km from the ground ({\em i.e} $\approx 0.3u_*/f$). With respect to the outer layer, the characteristic length scale of turbulence diminishes rapidly in the surface layer and in its lowest portion, such as the roughness sublayer and the urban canopy layer. Briefly, the depth of the `urban canopy layer' is about the mean height ($z=H$) of the roughness elements, and that of the roughness sublayer (RSL) is about $2$-$5$ times the mean height ($H$) of the roughness elements, where the effects of the roughness elements are perceived. The inertial sublayer is the region between the outer layer and roughness sublayer. It is a computationally challenging endeavour to capture the numerical details of turbulence simultaneously in the outer layer and in the surface layer. One of the strategies used in atmospheric boundary layer simulations is to employ a turbulence modelling scheme for the very lowest portion of the boundary layer (see Fig~\ref{fig:uabl}).  %

To predict turbulence accurately in the UABL, it is important to understand how the urban canopy layer~--~formed by the complexity of buildings -- alters the air flow in the surface layer due to the drag and shear forces exerted by the buildings. Moreover, buildings modify the net energy due to absorption of radiation, and alters the turbulent production in the surface layer, {\em e.g.} due to the formation of wake vorticies. A comprehensive knowledge of turbulence in the UABL can be gained by applying fast and accurate computational atmospheric modelling techniques, which is important {\em e.g.} for urban planning and disaster response, to name a few. The Reynolds averaged Navier-Stokes~(RANS) model employing a land-surface parameterization scheme~\cite[e.g.][]{Grimmond2011,Resler2017} defines a length scale for the surface roughness, and the stress exerted by the rough surface is computed assuming the Monin-Obukhov similarity theory in the inertial sublayer ({\em e.g.} see Fig~\ref{fig:uabl}). While the RANS approach is pragmatic for mesoscale modelling of the UABL, in order to sufficiently capture turbulent mixing and transport within the UABL, a large eddy simulation (LES) methodology with a robust model for the drag and shear forces of the roughness elements is the primary topic of the present article.
\subsection{Known challenges of turbulence modelling in the UABL}\label{sec:inta}
\begin{figure}[t]
  \centering
  \includegraphics[height=8cm]{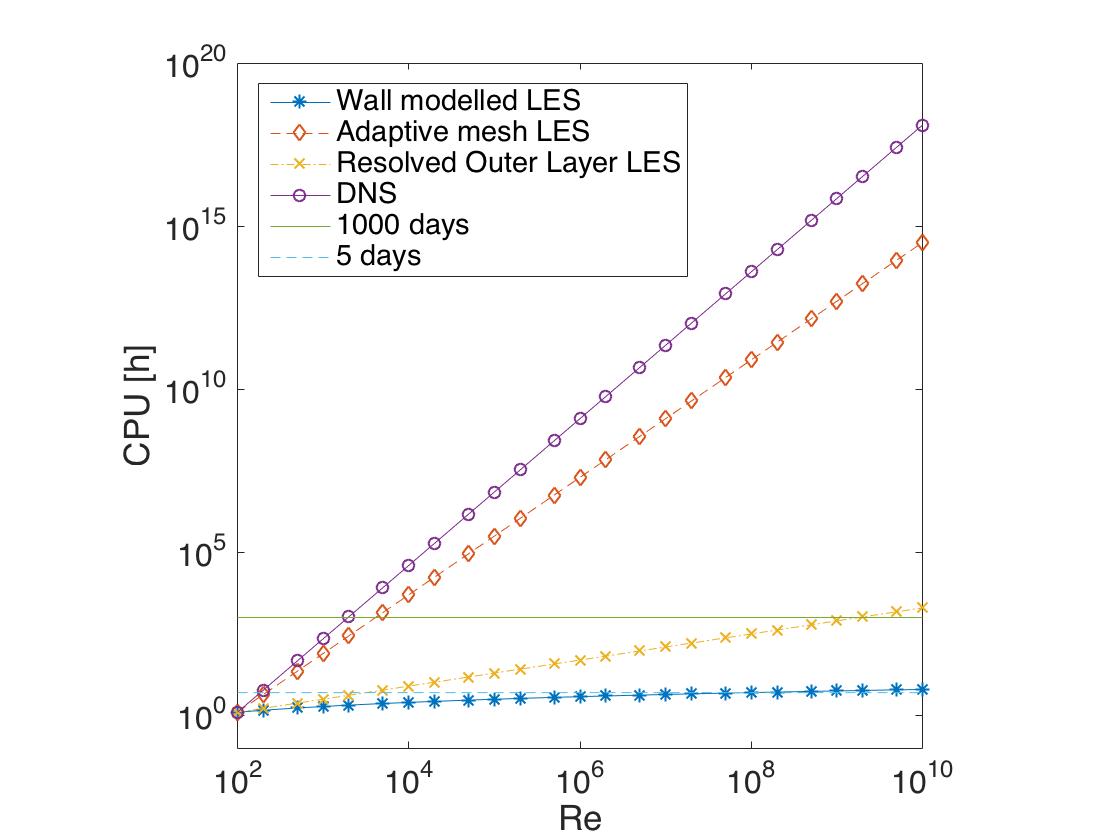}
  \caption{Comparison of computational cost of LES}
  \label{fig:cpules}
\end{figure}

LES is a promising computational methodology in which the  resolved fraction of turbulent structures is filtered by a length scale, $\Delta_{\hbox{\tiny{LES}}}$, and unresolved turbulent structures are represented by a subfilter scale~(SFS) model~\cite[][]{dear70,moeng}. However, the computational promise of LES breaks down in the atmospheric boundary layer~\cite[see][]{Chow2005}, as well as in a turbulent boundary layer over a flat plate~\cite[see][]{Chung2009} because the characteristic length scale of energetic turbulent structures ({\em i.e.} large eddies) diminishes rapidly in the surface layer, thereby requiring too fine a spatial and temporal resolution in the near-surface zone~\cite[][]{Senocak2007}. One of the strategies to address this issue is the `wall-resolved LES' that combines the benefits of LES with that of the adaptive mesh refinement~(AMR) technique~\cite[e.g.][]{Berger84,Alam2015} -- in which the resolution can be increased locally in the near-surface zone. The cost of wall-resolved LES scales like $\mathcal O(\mathcal Re^{1.8})$ as $\mathcal Re\rightarrow\infty$, which is too costly if Reynolds number $\mathcal Re$ is large. The other strategy is the `wall-modelled LES' that improves the turbulence prediction in the near-surface zone without primarily depending on mesh refinement techniques. An advantage of such a wall-modelled LES is that the resolution requirements relative to the boundary layer thickness depends weakly on the Reynolds number~\cite[][]{Pope2000,Chung2009}. It is worth mentioning that the computational cost for wall-modelled LES scales like $\mathcal O(\log\,\mathcal Re)$, which is significantly lower than that of wall-resolved LES, $\mathcal O(\mathcal Re^{1.8})$. An estimated computational cost for wall-resolved LES and that for wall-modelled LES with respect to a turbulent flow over flat plates is demonstrated in Fig~\ref{fig:cpules}, which is according to the cost per one time unit required on an AMD opteron cluster~\cite[e.g.][]{Piomelli2002}. As indicated by Fig~\ref{fig:cpules}, wall-resolved LES for the UABL is computationally challenging, which would lean heavily on massively parallel computing resources. The range of length and time scales of atmospheric turbulence is too wide for wall-resolved LES of UABL to be feasible, which suggests that advancing the wall-modelled LES methodology for UABL is an important aspect in the field of computational atmospheric modelling. 
\subsection{Recent work on wall-resolved and wall-modelled LES for UABL}
There exists a number of articles investigating the accuracy of wall-resolved LES. For example in~\cite{Castro2006}, the complexity of buildings and streets is addressed by clustering the computational nodes next to the surface of the buildings. The accuracy and computational benefits of the AMR approach to atmospheric boundary layer simulations was discussed by~\cite{Skamarock89}~\cite[see also][]{Skamarock93}. The AMR approach was also investigated for other geophysical flow simulations by~\cite{Jablonowski2006} and~\cite{Jablonowski2009}. A major drawback of AMR for atmospheric boundary layer simulation stems from the existence of mesh-refinement interfaces between one level of an adaptive mesh to the other. It was found that coupling LES with AMR often produces inaccurate results, for example, erroneous distribution of resolved energy in physical and spectral space was reported by~\cite{Goodfriend2015} and~\cite{Goodfriend2016}. The source of this error was discussed in more details by~\cite{Moeng2007}, \cite{Talbot2012}, and~\cite{Mirocha2013} in the context of LES in the Weather Research and Forecasting model. %
For LES of flow over an array of building-like obstacles, \cite{Philips2013} observed wiggles in the mean velocity and the Reynolds stress in regions where the profiles passes through the interface between two resolutions of the adaptive mesh. For simulating dispersion of pollutants in a London neighbourhood, \cite{Castro2009} considered an unstructured polyhedral mesh, achieving  a smooth transition from a fine mesh region to a coarse mesh region, which avoids the  grid refinement interface. For isothermal and dry air flow over a periodic array of surface mounted cubes, \cite{Goodfriend2016} reported that explicit filtering and reconstruction of SFS stress help mitigate turbulence modelling errors across mesh interfaces. In other words, designing wall-resolved LES based on AMR techniques must invent SFS modelling techniques adapted to  mesh refinement transition zones.

In contrast, wall-modelled LES for UABL suffers from the lack of a methodology dealing with urban roughness elements within the context of an existing wall-modelling scheme.
A principal idea of wall-modelling is to blend the LES filtering scale $\Delta_{\hbox{\tiny LES}}$ with the wall distance in order to better satisfy Kolmogorov's similarly hypothesis within the surface layer, where the cut-off scale for dissipating eddies is smaller than that in the outer layer. For instance, \cite{Mason92} suggested $1/(\Delta^*_{\hbox{\tiny LES}})^2 = 1/(\Delta_{\hbox{\tiny LES}})^2 + 1/(\kappa z)^2$ to obtain the cut-off scale $\Delta^*_{\hbox{\tiny LES}}$, where $\kappa$ is the von Karman constant. %
The wall-modelling method of~\cite{Mason92} was examined by~\cite{Senocak2007} for LES of a neutrally stratified atmospheric boundary layer over an aerodynamically rough surface,  where the results are compared  among three distinct wall-modelling approaches, as well as with respect to LES without any wall-modelling. Using wind tunnel measurements, \cite{Brown2001} validated a canopy stress method to model the effect of mountain-like obstacles without needing a boundary layer mesh around the obstacles~\cite[see also][]{Hao95}. \cite{Senocak2007} assumes that the interaction between energetic surface layer eddies has resulted into a stress that would otherwise be due to a canopy, and thus, examined the canopy stress method of~\cite{Brown2001} as a potential wall-modelling strategy in the atmospheric boundary layer. The results indicate that blending the cut-off scale by the method of~\cite{Mason92} is equivalent to treating the surface layer turbulence in the form of a canopy stress, albeit in the absence of a canopy layer. 

\subsection{Outline for present research}

There is thus a growing need for sophisticated near-wall modelling schemes for LES~\cite[][]{Senocak2007,Chung2009}. In the present article, we discuss a wall-modelling approach for LES of the UABL and examine its accuracy for resolving near-surface turbulence. In particular, we extend the canopy stress method of~\cite{Brown2001} to apply boundary conditions on the surface of the urban roughness elements, and validate the results against a classical CFD method resolving roughness elements. Without blending $\Delta_{\hbox{\tiny LES}}$ to the wall distance~\cite[e.g.][]{Mason92}, we consider blending the SFS stress to the length scale of surface-layer energetic eddies~\cite[e.g.][]{Nicoud99}. 

To formulate a canopy stress model for turbulence in the UABL, we consider a turbulent flow over a periodic array of surface mounted cubes. Using a periodic array of cubes to idealize turbulence in a modern city greatly helps understand the present LES results because the same flow was simulated by other researchers, which was also investigated by wind tunnel measurements. In Section~\ref{sec:meth}, we discuss how to couple the wall adaptive eddy viscosity with the canopy stress experienced by the roughness elements.  In Section~\ref{sec:num}, we present the physical model and summarize our numerical results. Finally, Section~\ref{sec:con} comments on the present findings, and discusses potential future extension of the LES methodology developed in this work.

\section{Methodology}\label{sec:meth}

\subsection{Large eddy simulation of UABL}
Let us consider an LES filtering approach that is similar to the method studied by~\cite{Finnigan2009} for canopy turbulence. Thus, energetic eddies are filtered with respect to a box that may be a multiply-connected airspace in the urban canopy layer, containing both air and solid~\cite[e.g.][]{Finnigan2009,Yan2017}. We then solve the following filtered Navier-Stokes equation ($u_i$ is used instead of $\langle u_i\rangle$ for simplicity)
\begin{equation}
  \label{eq:nse}
  \frac{\partial u_i}{\partial t} + u_j\frac{\partial u_i}{\partial x_j} = -\frac{1}{\rho_0}\left(\frac{\partial p}{\partial x_i} + \delta_{i1}\frac{\partial\langle P\rangle}{\partial x_1}\right) + \frac{\partial\tau_{ij}}{\partial x_j} + \frac{\partial\tau^c_{ij}}{\partial x_j}, 
\end{equation}
\begin{equation}
  \label{eq:inc}
  \hbox{and}\quad\frac{\partial u_i}{\partial x_i} = 0.
\end{equation}
Here, the unfiltered velocity $\tilde u_i = u_i + u_i'$ is the sum of a filtered part $u_i$ and a sub-filter scale part $u'_i$. The SFS stress $\tau_{ij}=\langle u_i\rangle\langle u_j\rangle - \langle u_iu_j\rangle$ (force per unit area divided by density) is due to the LES filtering, and the canopy stress $\tau^c_{ij}$ is due to the presence of urban roughness elements.
In the outer layer (see Fig~\ref{fig:uabl}), the resolved eddies contain most of the turbulent kinetic energy, and provide the length and time scale to model the effects of the rest of the small-scale turbulence through the SFS stress tensor $\tau_{ij}$~\cite[][]{dear70,Senocak2007,Chung2009}. In the outer layer, the deviatoric part of the SFS stress $\tau_{ij}$ is related to the filtered rate of strain $\mathcal S_{ij}$ such that~\cite[][]{smagorinsky} 
$$\tau_{ij} = 2\nu_{\tau} \mathcal S_{ij} - \frac{1}{3}\tau_{kk}\delta_{ij}\quad\hbox{and}$$
$$\nu_{\tau} = (\Delta_{\hbox{\tiny LES}})^2\sqrt{2\mathcal S_{ij}\mathcal S_{ij}},$$
where $\Delta_{\hbox{\tiny LES}} = C_s\sqrt[3]{\Delta x\Delta y\Delta z}$ is the filter width, $C_s$ the Smagorinsky coefficient, and $\Delta x$, $\Delta y$, and $\Delta z$ denote of the size of a computational cell in $x$, $y$ , and $z$ directions respectively. After the pioneering work had been done by~\cite{smagorinsky}, the success of the above scheme for the SFS stress was witnessed for homogeneous isotropic turbulence, shear layer turbulence, wake turbulence, and turbulent jets. 

In the presence of an approximately flat and aerodynamically rough surface ({\em e.g.} earth's surface), there exists a viscous layer in proximity of the surface at $z < 5\nu/u_*$, where $\nu$ is kinematic viscosity and the friction velocity is given by $u_*=\sqrt{(\tau_{31})^2 + (\tau_{32})^2}$. Clearly, the thickness of the viscous layer is too small to be resolved. \cite{dear70b} and \cite{moeng} suggested to bypass the viscous layer through a bottom boundary condition, where horizontal components of the turbulent stress tensor are estimated by fitting a logarithm law of a rough wall such that 
$$
\tau_{3i} = -u_*^2\frac{u_i}{|\bm u|} = -\left(\frac{\kappa}{\ln\frac{z_1}{z_0}}\right)^2|\bm u(z_1)|u_i(z_1).
$$
Here, $z_0$ is the roughness length scale and $z_1$ is the first vertical grid point above the ground.

The immediate effects of earth's surface include increasing turbulence anisotropy and decreasing the cut-off scale for energetic eddies. Thus, a blending of the turbulent eddy viscosity $\nu_{\tau}$ was suggested by~\cite{VanDriest56} so that the filter width of LES can be switched off to a mixing length model near the surface. For atmospheric boundary layer, one way of blending the filter width is to define the turbulent viscosity by
$$\nu_{\tau} = [(1-\exp(-z/h))^2\Delta_{\hbox{\tiny LES}}^2 + \exp(-z/h)^2(\kappa z)^2]\sqrt{\mathcal S_{ij}\mathcal S_{ij}}.$$
Considering the above blending function along with the dynamic Smagorinsky model in the outer layer,  \cite{Senocak2007} demonstrated that the blending approach produced the law of the wall slightly better than the method proposed by~\cite{Mason92}.

\subsection{Wall adaptive local eddy viscosity (WALE) model}\label{sec:cstm}
The basic idea of wall-modelling approach in LES~\cite[e.g. see][]{Chung2009} is to derive the SFS stress and the wall condition in a way that the numerical resolution in the near wall region remains independent of the Reynolds number $\mathcal Re$ or depends at most weakly on $\mathcal Re$~\cite[][]{Pope2000}. 
If the SFS stress were obtained by the classical model of~\cite{smagorinsky}, numerical experiments of LES for turbulent boundary layer over a flat surface indicate that the eddy viscosity overshoots the shear production with respect to SFS dissipation. Perhaps, it is because $\nu_{\tau}$ is obtained from the strain rate of the resolved `large eddies' that exist only in the outer layer, but not in the near-wall region. \cite{Nicoud99} argue that a proper near-wall scaling of SFS dissipation can be obtained by relating the SFS dissipation with both the strain rate and the rotation rate. For smooth surfaces such as in a channel flow, this approach was found effective for adjusting the eddy viscosity in the near wall region to switch off the SFS stress to a near-wall scaling such that $\nu_{\tau}$ behaves like $\mathcal O(z^3)$ as $z\rightarrow 0$. Note that the blending approach of~\cite{VanDriest56} switches off $\nu_{\tau}$ to the well known mixing length model, {\em i.e.} $\mathcal O(z)$. 
To mimic the energy transfer, {\em e.g.} from the resolved scales to the sub-filter scales, the WALE model~\cite[][]{Nicoud99} defines the eddy-viscosity %
$$
\nu_{\tau} = \left(\Delta_{\hbox{\tiny LES}}\right)^2\frac{(\mathcal S_{ij}^d\mathcal S_{ij}^d)^{3/2}}{(\mathcal S_{ij}^d\mathcal S_{ij}^d)^{5/4} + (\mathcal S_{ij}\mathcal S_{ij})^{5/2}},
$$
where we engage the velocity gradient tensor such that
$$
\mathcal S_{ij}^d = \frac{1}{2}\left[\left(\frac{\partial u_i}{\partial x_j}\right)^2+\left(\frac{\partial u_j}{\partial x_i}\right)^2\right] - \frac{1}{3}\delta_{ij}\left(\frac{\partial u_k}{\partial x_k}\right)^2.
$$
Various numerical tests of the WALE model (primarily for pipe flow and channel flow) suggest a typical value of $C_s=0.325$, where the filter width is given by $\Delta_{\hbox{\tiny LES}} = C_s\sqrt[3]{\Delta x\Delta y\Delta z}$.
\begin{figure}
  \centering
  \includegraphics[height=8cm]{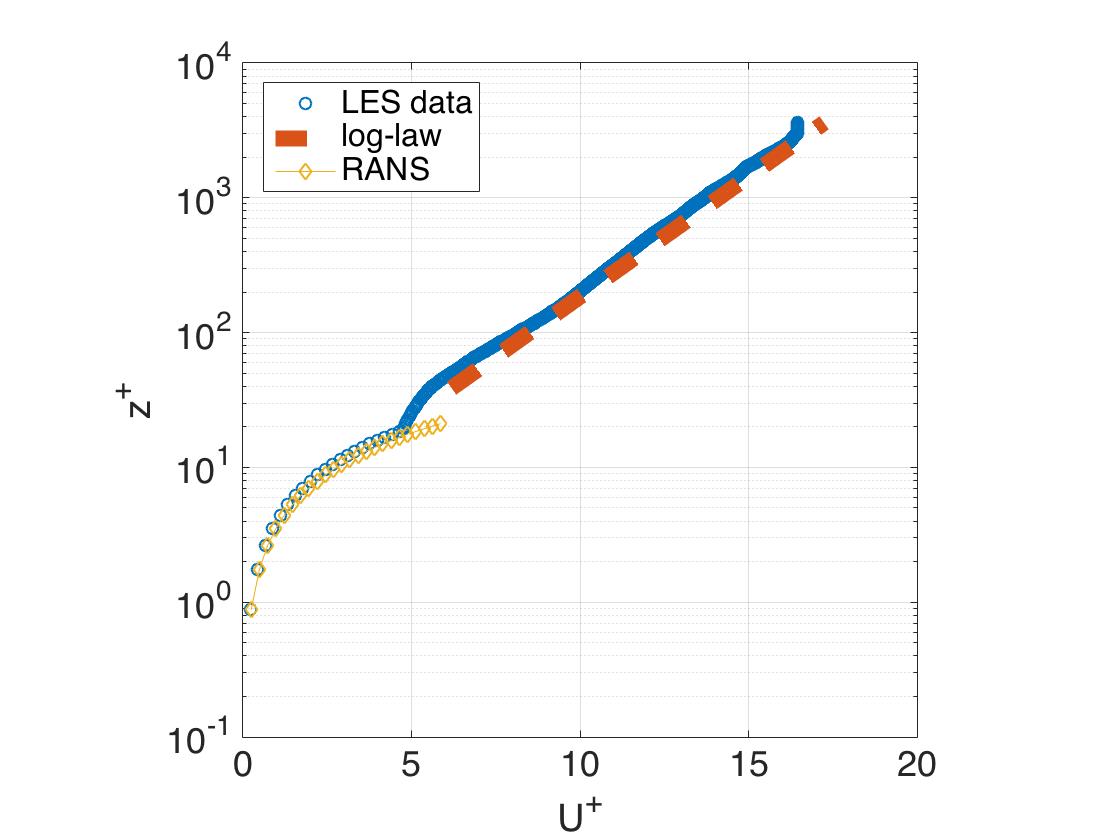}
  \caption{Predicting the atmospheric surface layer using WALE method}
  \label{fig:wale}
\end{figure}

To the best of authors' knowledge, the performance of the WALE model was not fully exploited in the literature dealing with atmospheric boundary layer simulations. Here, we are interested to demonstrate an overall idea of the WALE model for a neutral atmospheric boundary layer flow over an aerodynamically rough wall. Consider a computational domain of $3~\hbox{km}\times 3~\hbox{km}\times 1~\hbox{km}$ with a spatial resolution of $128\times 128\times 192$, where the flow is driven by a mean geostrophic wind of $\langle 10, 0,0\rangle$~m/s. Using a stress condition on the bottom boundary along with the WALE model described above, the simulation is tested for $24$ hours of model time. As it is demonstrated in Fig~\ref{fig:wale}, the LES-WALE prediction of the mean wind $U(z)$ indicates a very good agreement with the logarithmic law of the wall for $z^+ > 30$, where the vertical resolution is $\Delta z^+ = 18.75$ in wall units $\nu/u_*$. One notices for the viscous sublayer $z^+ < 30$ that the WALE model switches off automatically to the RANS model predicting a velocity that varies linearly with the wall distance, {\em i.e.} $U(z) = z$. It is worth mentioning that the simulation geometry and the boundary conditions are equivalent to what is reported by~\cite{Senocak2007}. However, the first grid point is located $2$~m from the surface in the work of~\cite{Senocak2007}. In our case, it is $5.2$~m away from the surface. Unlike~\cite{Senocak2007}, neither the dynamic Smagorinsky model nor a bonafide wall-modelling scheme was used in our simulation, except the WALE model. However, in comparison to the results reported by~\cite{Senocak2007} ({\em e.g.} Figs3,5 therein), the present LES-WALE prediction of mean vertical wind speed is sufficiently accurate with respect to the law of the wall. The rest of the article demonstrates the accuracy of the canopy stress method to simulate turbulence in the UABL. 
\subsection{Canopy stress parameterization scheme}
The conventional Computational Fluid Dynamics (CFD) approach to simulate fluid flow past an urban roughness element would employ a structured or unstructured mesh that conform to the solid body~\cite[e.g.][]{Castro2006,Goodfriend2016}. In such a conventional approach, the volume occupied by the fluid is discretized into a mesh conforming to the solid body. Thus for LES, the filtered momentum equations need be transformed to a curvilinear coordinate system aligned with the mesh. 
Instead of employing a conventional CFD mesh around urban roughness obstacles~\cite[e.g.][]{Brown2001}, the idea of the canopy stress method for the UABL is to consider a structured mesh that overlaps with the obstacles~\cite[see][]{Hao95,Brown2001}. As a result, some computational cells may contain both solid and fluid, and some may contain solid only. The same is true for the filtering process. %
Using a box filter, differentiation and LES filtering do not commute for linear pressure gradient and viscous terms~\cite[][]{Finnigan2009}, which leads to an additional stress $\tau^c_{ij}$ term in the filtered momentum equations. This additional term accounts for the skin friction and  the pressure loss experienced by an eddy passing over the obstacle~\cite[e.g.][]{Hao95,Brown2001,Coceal2004,Finnigan2009,Yan2017}. In a formal statement, if $k_0$ isolated obstacles interact with the fluid flow, the vertical gradient of such a stress $\partial\tau^c_{i3}/\partial z$ may be expressed by~\cite[see][]{DeLemos2006}
$$
\frac{\partial\tau^c_{i3}}{\partial z} \equiv f_i = \frac{\nu}{\Delta V}\iint_{S_k}\frac{\partial u_i}{\partial n}dS - \frac{1}{\Delta V}\sum_{k=1}^{k=k_o}\iint_{S_k}pn_idS,
$$
where $S_k$ denotes the surface of the $k$-th obstacle and on the right hand side, the first term estimates the skin friction and the second term estimates the pressure drop. %

Assuming that the effect of $\tau^c_{ij}$ vanishes further away from the ground, \cite{Brown2001} considered the pressure loss only, {\em i.e.} $\partial\tau^c_{i3}/\partial z = -C_d\cos^3\left(\frac{\pi z}{2h_c}\right)|\bm u|u_i$ for $z \le h_c$, where $C_d$ is a drag coefficient in m$^{-1}$ units. The prediction of flow over a ridge by this formulation of the canopy stress was found in agreement with the wind tunnel data~\cite[][]{Brown2001}. A similar formulation, such as  $f_i = C_d A |\bm u| u_i$ was adopted by~\cite{Hao95} for boundary layer flow over isolated roughness elements.
In the present article, we follow the numerical results of turbulent flow in porous media presented by~\cite{DeLemos2006} and treat the computational cells overlapping the region of roughness elements as a `porous zone'. A benefit of this assumption is that the canopy forcing term in the LES equations~(\ref{eq:nse}) can be parameterized with respect to a length scale $d$ and a dimensionless number $\phi$. Thus, we consider the following expression
$$
f_i = \frac{\nu}{\mathcal K}u_i + \frac{C_f|\bm u|}{\sqrt{\mathcal K}}u_i,
$$
where
$$
\mathcal K = \frac{d^2\phi^3}{(1-\phi)^2},\quad \mathcal F = \frac{1.75d}{150(1-\phi)},\quad\hbox{and}\quad C_f = \frac{\mathcal F}{\sqrt{\mathcal K}}.
$$
If the linear term $\nu u_i/\mathcal K$ is omitted, then the expression for $f_i$ takes the equivalent form used, for example, by \cite{Hao95}, \cite{Coceal2004}, and \cite{Finnigan2009}. Note that all wave numbers larger than $2\pi/\Delta_{\hbox{\tiny LES}}$ are filtered by LES. The `porous zone' serves to filter wave numbers larger than $2\pi/d$~\cite[see][]{Hao95}. %

For example, \cite{Branford2011} considered the canopy stress model in direct numerical simulation~(DNS) at roughness Reynolds number $\mathcal Re_{\tau}=500$. In their work, turbulence over an array of cubical obstacles of height $H$ is modelled directly based on the viscous stress and the canopy stress. A number of other large eddy simulations using the canopy model have been performed for flow within and above a vegetative canopy~\cite[][]{Finnigan2009,Bailey2013} or a forest like canopy~\cite[][]{Shaw92}, flow over an array of cubes~\cite[][]{Coceal2004}, and flow over mountain-like obstacles~\cite[][]{Alam2011,Liu2016}. \cite{Brown2001} found that a canopy model would simulate the near wall flow condition in an LES, where the mesh is insufficient to capture the viscous layer. \cite{Senocak2007} simulated a neutrally stratified atmospheric boundary layer using LES to investigate the role of canopy models to represent the near-surface flow. Time series of friction velocity and the span-wise component of the instantaneous vorticity are in good agreement between the LES/canopy model and the hybrid RANS/LES model employed by~\cite{Senocak2007}. In other words, the surface stresses experienced by building-like obstacles can also be estimated in a similar way that the canopy stress model was derived.

\section{Results and discussion}\label{sec:num}
\subsection{Flow through an infinite array of surface mounted cubes.}
The simulation considered in this section follows the work of~\cite{Goodfriend2016}, which represents a periodic array of building-like obstacles. Periodic boundary conditions are employed in both the stream-wise and the span-wise directions with respect to a single cube, representing an infinite array of obstacles with a height of $H=150$~[m] and a size of $150\times 150\times 150$~[m$^3$] unless it is stated otherwise.

The computational domain is $L_x\times L_y\times L_z = 600\times 600\times 510$~[m$^3$], where a cube is placed at the center of the domain, leading to a distance of $450$~[m] between two cubes in the infinite array. All simulations have considered a coarse mesh of $64\times 64\times 48$ cells and a fine mesh of $128\times 128\times 96$ cells. In the coarse mesh, the vertical resolution is $\Delta z_{\min}=1.6$~[m] adjacent to the ground, which increases gradually to $\Delta z_{\max}=16$~[m] and remains constant for the upper half of the domain. In the fine mesh, the grid spacing is reduced by a factor of $2$ in each direction, having $32$ points per side of the obstacle. Note that this grid spacing is about the same (or relatively fine) with respect to previous work, e.g.~\cite{Tseng2006,Hsieh2009,Salim2011,Goodfriend2016}.  %
The flow is driven in the stream-wise direction by a pressure gradient that is adjusted at each time step so that the bulk velocity $U_b = \frac{1}{L_z}\int_0^{L_z}\langle u\rangle_{xy} dz$ is approximately $2.53$~m/s. The Reynolds number of the simulated flow is $\mathcal Re=U_bH/\nu\approx 2\times 10^7$.

At $t=0$ the mean stream-wise velocity is assigned a surface layer logarithmic profile with a surface roughness $z_0$~[m] and a friction velocity $u_*$~[m/s], where the wall-normal and the span-wise velocities are assigned to zero. A perturbation is added  to this mean velocity near the ground so that the steam-wise perturbation is sinusoidal in the span-wise direction, and the span-wise perturbation is sinusoidal in the stream-wise direction; {\em i.e.}
$$
u(x,z) = \frac{u_*}{\kappa}\ln\left(\frac{z}{z_0}\right) + ze^{-\sigma z^2}\sin(\beta y)(1+{\tt R_{20}})
$$
and
$$
v(x,z) = ze^{-\sigma z^2}\sin(\alpha x)(1+{\tt R_{20}}).
$$
Here, ${\tt R_{20}}$ represents a random number that is about $\pm 20$\% of the applied perturbation, which aims to break the symmetry; a value of $\sigma$ is chosen to attain the perturbation in a near-ground region; $\alpha$ and $\beta$ are wave numbers of the perturbation, and $\kappa$ is the von Karman constant.  
\subsubsection{Wind tunnel measurement at $\mathcal Re=3\,854$}
The above setting of the simulation extends a wind tunnel experiment of channel flow past an array of $250$ cubes arranged in $25$ rows and $10$ columns~\cite[][]{Meinders99}. The size of each cube was $H=15$~mm and the array was spaced equidistantly at $45$~mm (face-to-face) in both the stream-wise and span-wise direction. In the wind tunnel experiment, measurements were conducted around a cube in the $18$-th row, near the centerline of the array. The bulk velocity was estimated $U_b=3.8$~m/s based on the measured mass flow rate of $6.85\times 10^{-3}$~kg/s, leading to a Reynolds number of $\mathcal Re=3\,854$.

\cite{Goodfriend2016} simulated a channel flow at $\mathcal Re\approx 3\,800$ using a parameter setting that is dynamically equivalent to the channel flow presented by~\cite{Meinders99}. However, in the LES of~\cite{Goodfriend2016}, $U_b=2.53$~m/s, $H=150$~m, and  a constant eddy viscosity of $\nu=0.1~\hbox{m}^2$/s were used. Using a locally refined mesh conforming to the obstacle, the LES prediction of mean quantities along five vertical lines aligned in the stream-wise direction exhibited an excellent agreement with the wind tunnel measurements~\cite[][]{Goodfriend2016}. 
\subsubsection{CFD simulation with a conforming mesh} 
\begin{figure}
  \centering
  \begin{tabular}{cc}
    \includegraphics[height=6cm]{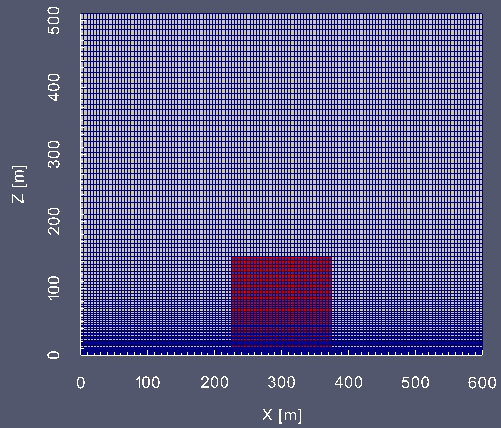}&
    \includegraphics[height=6cm]{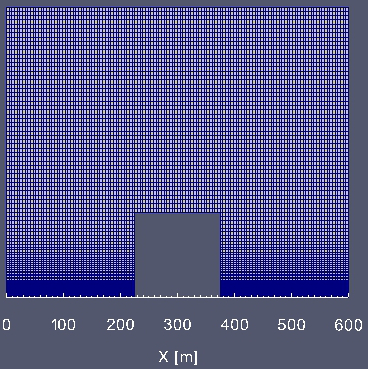}\\
    $(a)$ & $(b)$\\
  \end{tabular}
  \caption{Vertical slices of two meshes: $(a)$ the mesh for stress model, where the cells occupied by the obstacle is marked by red colour, and $(b)$ the mesh for CFD model, where the is generated around the obstacle.}
  \label{fig:msh}
\end{figure}

To gain a quantitative comparison for the present simulation at $\mathcal Re=2\times 20^7$, we consider a classical CFD method based on a mesh conforming to the the solid body. In other words, we remove the canopy stress term from Eq~(\ref{eq:nse}) and apply the no-slip boundary conditions on the surface of the obstacles. This allows us to directly compare the LES prediction using the canopy stress method with that using the CFD method. In~\cite{Chung2009}, LES prediction of mean wall normal velocity profiles were compared with respect to five Reynolds number in the range of $[2\times 10^3,2\times 10^7]$, and all of the five mean profiles expressed with the `wall unit' collapsed in a self-similar manner. We have used the low $\mathcal Re$ simulations of~\cite{Meinders99} and~\cite{Goodfriend2016} to gain a qualitative understanding of our simulation, and a comparison with CFD method provides a quantitative measure of accuracy in our LES-canopy prediction.  Fig~\ref{fig:msh} demonstrates a vertical cross-section of the mesh used for both approaches.  %
\subsection{Primary validation result}
\subsubsection{Coherent structures}
As the mean boundary layer flow $(U(z),0,0)$ approaches a surface mounted obstacle, its top and side edges cause flow separation because a span-wise wind shear $\partial v/\partial y$ is caused by the SGS stress component $\tau_{22}$, leading to a downstream wake of vortex re-attachment~\cite[][]{Hsieh2009,Goodfriend2016}, which is illustrated schematically in Fig~3 of~\cite{Meinders99}. Such an overall description may also be given by the vorticity equation. Ignoring the viscous effects, the mean vorticity $\langle 0,\omega_2,0\rangle$  evolves according to
$$
\frac{\partial\omega_2}{\partial t} + u_j\frac{\partial\omega_2}{\partial x_j} = \omega_2\frac{\partial v}{\partial y},
$$
where the vortex stretching term on the right-hand side indicates that the streamlines pass around the obstacle. Thus, the strength $\omega_2$ of the mean span-wise vortex tube increases as it approaches the obstacle. For example, one may notice that snow is scoped out in front of a tree or pole during a snow storm, which is also due to the increasing strength of the span-wise vortex tube. A possible net effect of strengthening the span-wise mean vortex-tube in the turbulent flow around a building-like obstacle is the generation of the coherent flow structures~\cite[e.g.][]{Castro2006,Goodfriend2016}. 

To investigate the turbulent exchange of momentum by coherent structures in the vicinity of building-like roughness elements, we consider the unorganized random background flow initialized at $t=0$ and its  temporal evolution predicted by the canopy stress model, and the result is compared with that predicted by the CFD model. We consider two methods for identifying coherent structures, where  the rotation tensor $\displaystyle r_{ij} = \frac{\partial u_i}{\partial x_j} - \frac{\partial u_j}{\partial x_i}$ provides the vorticity field, and the `Q-criterion' provides the relative dominance of the rotation rate over the strain rate such that~\cite[see][]{Hunt88}
$$
Q = \frac{1}{2}\left[\left(\frac{\partial u_i}{\partial x_j} - \frac{\partial u_i}{\partial x_j}\right)^2 - \left(\frac{\partial u_i}{\partial x_j} + \frac{\partial u_i}{\partial x_j}\right)^2\right].
$$
The region with $Q>0$ indicates the rotation of turbulent eddies, and that with $Q<0$ indicates the deformation of turbulent eddies.  
\begin{figure}
  \centering
  \begin{tabular}{cc}
    Q-criterion  & Vorticity \\
    \includegraphics[height=4cm]{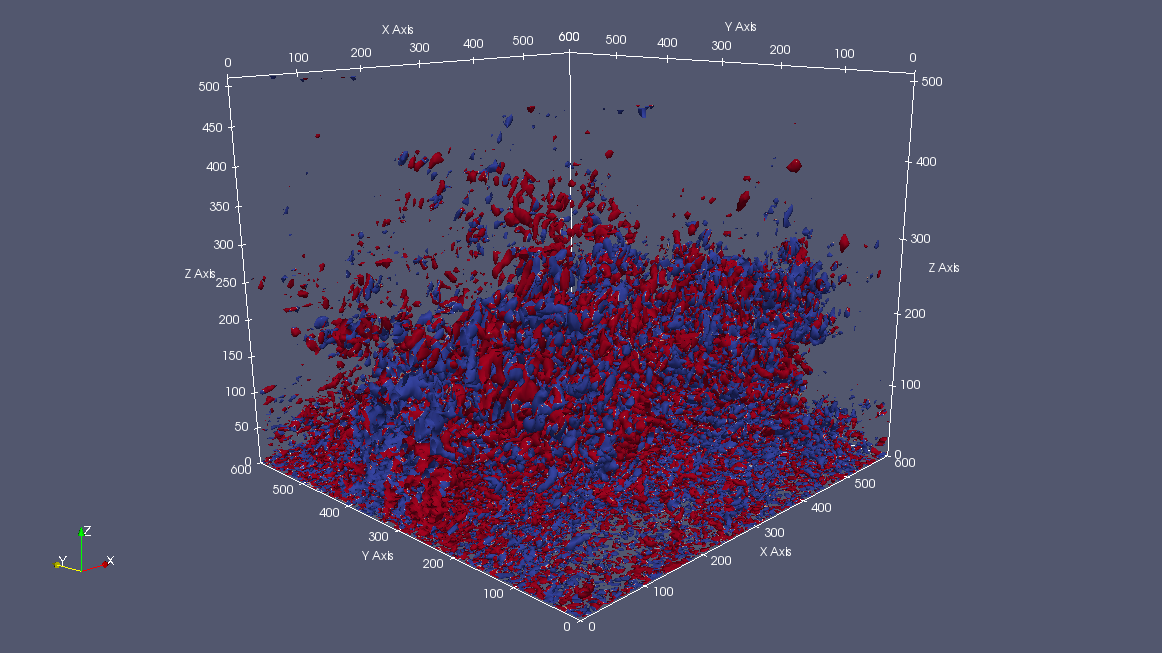}&
    \includegraphics[height=4cm]{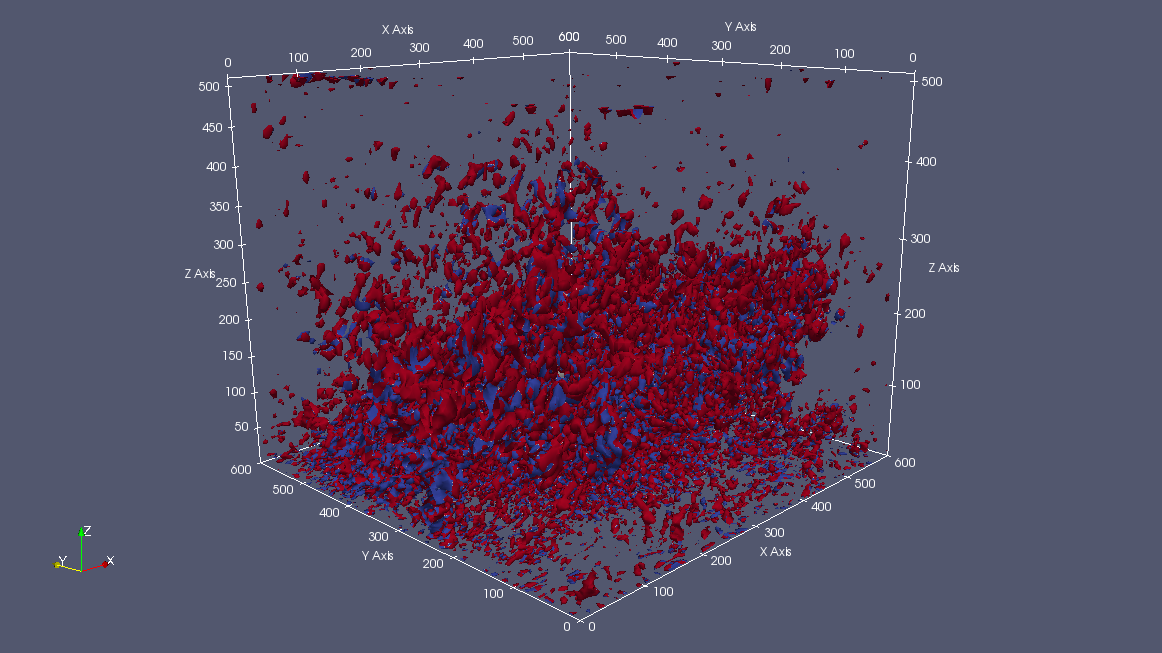}\\
    $(a)$ LES-WALE, Canopy & $(b)$ LES-WALE, Canopy\\
    \includegraphics[height=4cm]{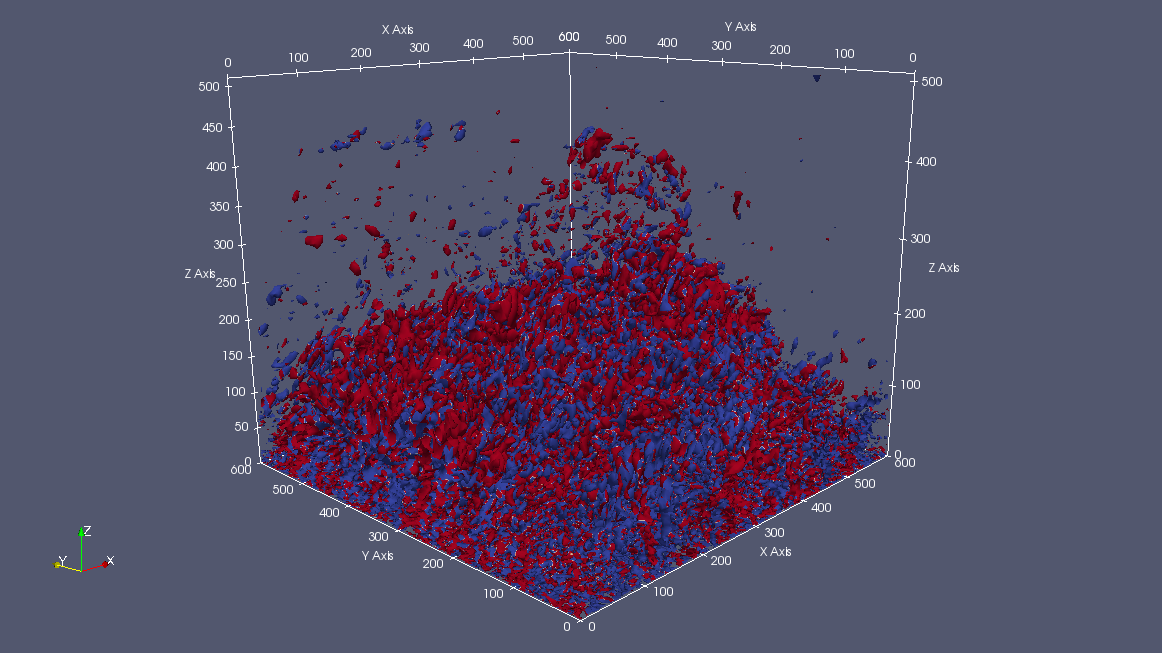}&
    \includegraphics[height=4cm]{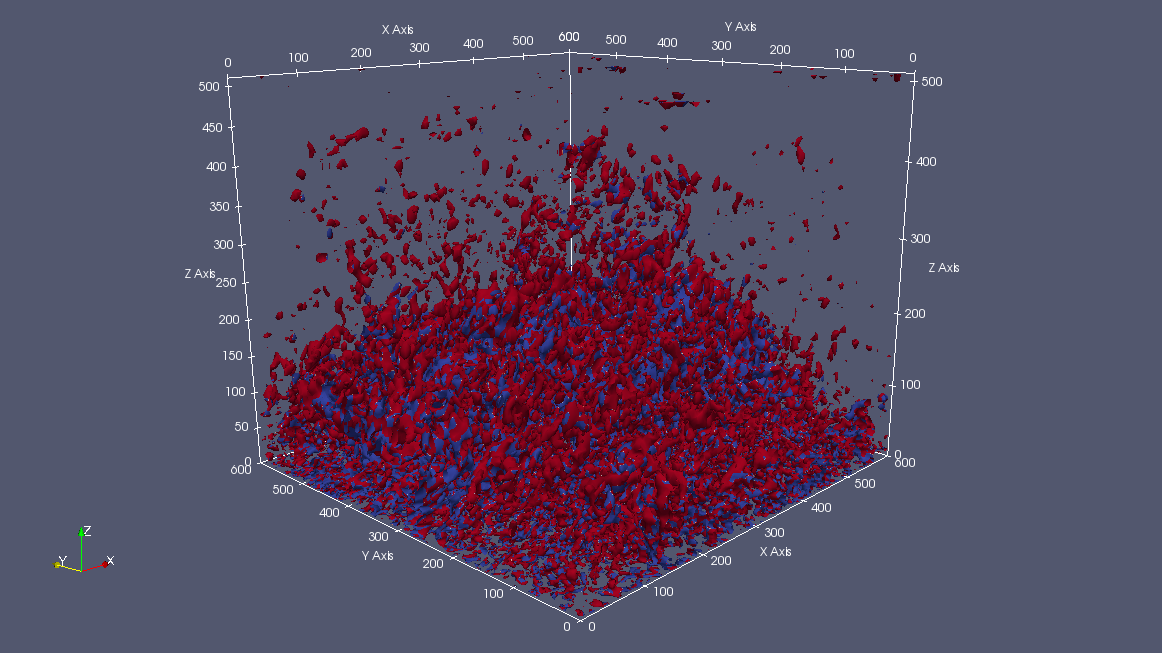}\\
    $(c)$ LES-WALE, CFD & $(d)$ LES-WALE, CFD\\
    \includegraphics[height=4cm]{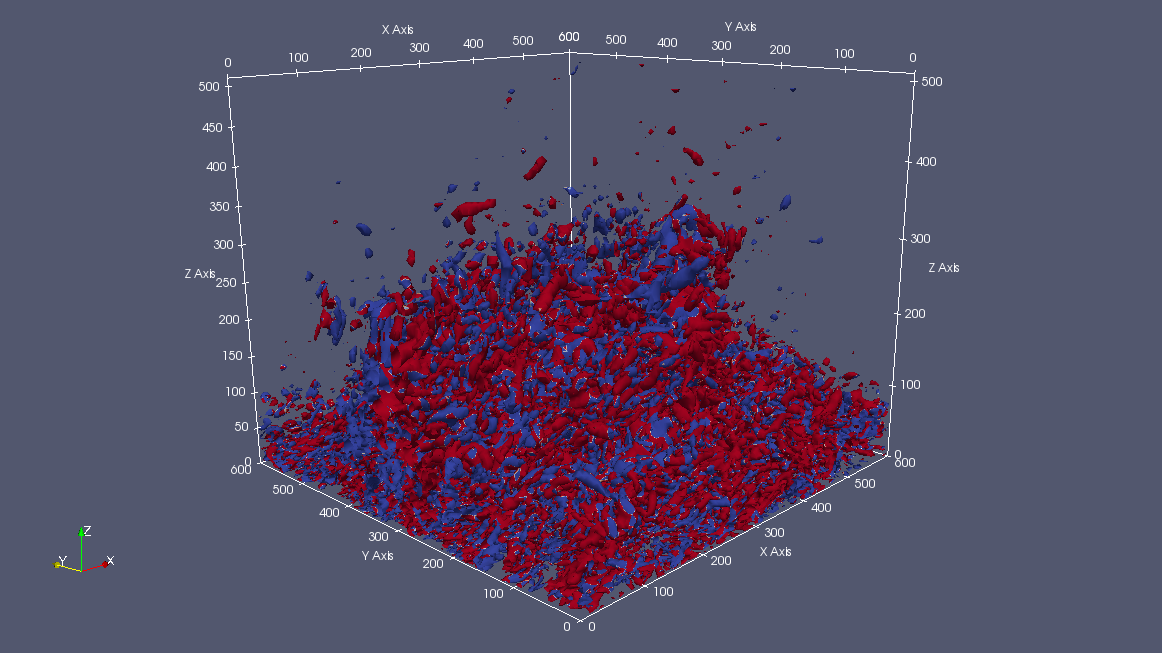}&
    \includegraphics[height=4cm]{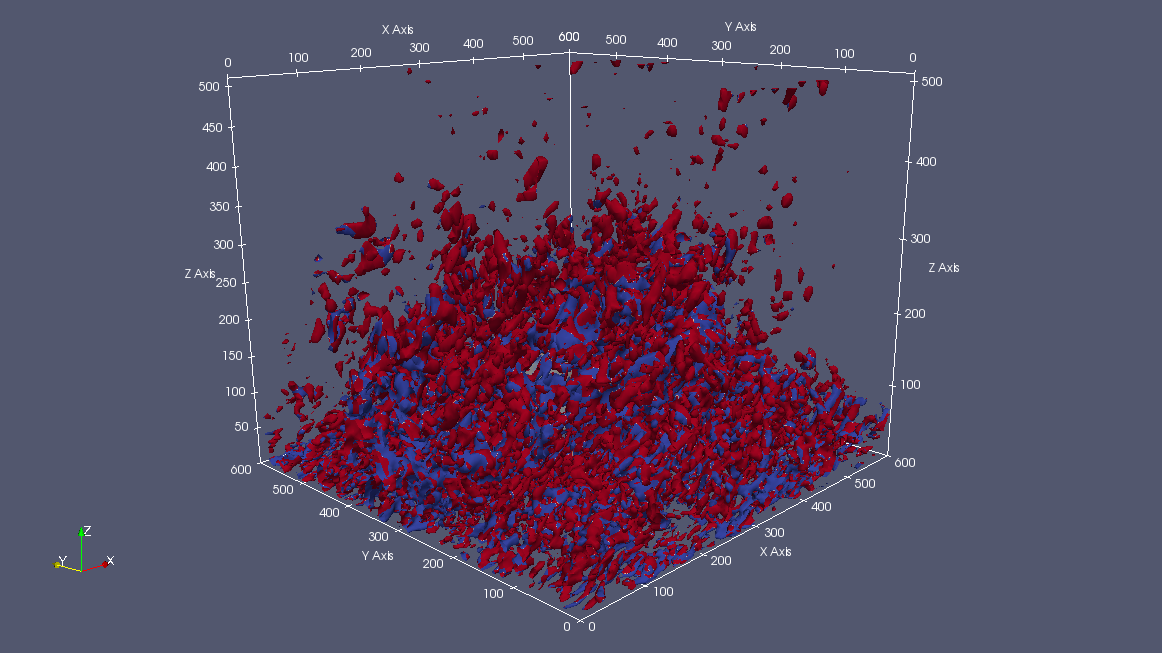}\\
    $(e)$ LES-Smagorinsky, Canopy & $(f)$ LES-Smagorinsky, Canopy\\
  \end{tabular}
  \caption{Left column: isosurfaces of the Q-criterion for values of -0.001 (blue) and 0.001 (red), and right column: isosurfaces of the vertical vorticity for values of -0.002 (blue) and  0.002 (red). Top row: LES-WALE using the canopy stress method; middle row: LES-WALE using the conforming mesh CFD method; bottom row: LES-Smagorinsky using the canopy stress method.}
  \label{fig:Q}
\end{figure}

In Fig.~\ref{fig:Q}, Q-criterion and vertical vorticity ($\omega_3$) are illustrated in first and second columns, respectively.  In comparison between the canopy stress model~({\em e.g.} Fig~\ref{fig:Q}$a,b$) and the CFD model~({\em e.g.} Fig~\ref{fig:Q}$c,d$), the range of turbulence scales resolved by the canopy stress model -- without generating a conforming mesh to resolve the solid surface -- is as wide as in the simulation by the CFD model using a  block-structured mesh conforming to the solid surface. Fig~\ref{fig:Q}$e,f$ also demonstrate Q-criterion and vertical vorticity computed with the canopy stress model, where the SGS stress was estimated by the classical Smagorinsky method. The success of the canopy stress model for resolving the near surface eddies can be seen by the equivalent richness of the flow structures predicted by the canopy model and the CFD model. The results also suggest that the need of mesh generation for a fluid flow around a solid body is paid off by the canopy stress model, which is intuitively cost effective. %
In an urban ABL, as the surface is approached, the scale of energetic eddies diminishes rapidly and turbulence becomes increasingly anisotropic. %
In the roughness sublayer, the present LES employs the wall-adapting local eddy viscosity model to approximate a portion of the SGS stress due to the unresolved small scale fluctuations, where the other portion of the SGS stress -- that is due to the interaction of roughness elements with the eddies around -- is approximated using the canopy stress model.
In contrast, the no-slip boundary conditions are directly applied by the CFD model. The comparison of coherent structures in Fig~\ref{fig:Q} indicates that the boundary conditions on the roughness elements are achieved through the canopy stress. The results do not indicate any additional numerical artifact of damping high frequency modes of turbulence. %
\subsubsection{Wall shear stress}
\begin{table}
  \centering
  \begin{tabular}{|ll|}
    \hline
    label & co-ordinate\\
    \hline
    UP1 & $(300,\,300,\,150)$\\
    UP2 & $(300,\,300,\,225)$\\
    UP3 & $(380,\,300,\,150)$\\
    UP4 & $(112.5,\,300,\,0)$\\
    UP5 & $(487.5,\,300,\,0)$\\
    UP6 & $(487.5,\,300,\,150)$\\
    \hline
  \end{tabular}
  \caption{Co-ordinates of six locations around the obstacle, where Reynolds stress and other quantities were recorded at each time step.}
  \label{tab:ups}
\end{table}
Wall shear-stress is an important parameter in the UABL, which is a primary cause of drag experienced by roughness elements in the urban canopy layer. We have post-processed the shear-stress experienced by the obstacle from the Reynolds stress ($\tau'_{ij}$), and denoted it by $u_*$ in Table~\ref{tab:bct}. The temporal evolution of $\tau'_{ij}$ was recorded at six locations around the obstacle (see Table~\ref{tab:ups}), and the shear stress was calculated as $u_*=(1/2)\sqrt{(\tau'_{13})^2 + (\tau'_{23})^2}$, resulting in six sets of time series data. First, the average over six locations was calculated, and then the average with respect to the last five hours of the time interval was obtained.

Similarly, minimum and maximum values of $\tau'_{13}$ and $\tau'_{23}$ on the ground were recorded at each time step from which the time series for the wall-friction velocity $u_{\tau}$ was obtained. The temporal average of $u_{\tau}$ in the last five hours of each simulation is reported in Table~\ref{tab:bct}. An excellent agreement of the results obtained by the canopy method and the CFD method was observed.
\begin{table}[h]
  \centering
  \begin{tabular}{ccccl}
    \hline
    No. of Cells             & $H$~[m]  & $u_*$~[m/s] & $u_{\tau}$~[m/s] & Turbulence closure\\
    \hline\\
    $128\times 128\times 96$ & $150$    & $0.0357$    & $0.0891$ &  LES-WALE + canopy\\
    $128\times 128\times 96$ & $75$     & $0.0124$    & $0.0704$ &  LES-WALE + canopy\\
    $128\times 128\times 96$ & $150$    & $0.0354$    & $0.0804$ &  LES-Smagorinsky, canopy\\
    $128\times 128\times 96$ & $150$    & $0.0441$    & $0.0799$ &  LES-WALE, CFD\\
    $56\times 56\times   64$ & $150$    & & &Goodfriend et. al. 2016\\
    \hline\\
  \end{tabular}
  \caption{The friction velocity $u_*$ experienced by the obstacle and the wall-friction velocity $u_{\tau}$.}
  \label{tab:bct}
\end{table}

\subsubsection{Intermittent burst of coherent structures}

Intermittency and bursting behaviour is a fundamental feature of atmospheric turbulence in the surface layer~\cite[][]{Mahrt99,Balsley2004,Costa2011,Alam2015}. In any turbulent flow, the dissipation of turbulent kinetic energy is usually confined into small sub-regions of individual eddies leading to a ``spotty'' appearance of coherent structures -- known as small-scale intermittency of turbulence~\cite[][]{Batchelor49,Frisch78}. Atmospheric turbulence is characterized by length and time scales that range over many orders of magnitude, where turbulence can also be intermittent due to organization of coherent structures on scales larger than the characteristic scale  of large eddies. This behaviour of atmospheric turbulence leads to episodic development of turbulence, which is called large-scale intermittency~\cite[][]{Mollo73,Balsley2004,Ansorge2014} or global intermittency~\cite[][]{Mahrt99,Costa2011}. The genesis of episodic bursting events in atmospheric turbulence is a direct result of downscale energy cascade of these large eddies~\cite[][]{Mollo73}. Consequently, work done in the region of a burst also leads to a burst of stress, which plays an important role in the transport of momentum and energy by stretching the region of small-scale turbulence.

In other words, features of large-scale intermittency in the surface layer include episodes of turbulent and calm periods in time series of the turbulence stress. For example, measurements of friction velocity for the night of 26/27 January, 2001 at a deforested site of Amazonian region display a chaotic organization of bursting events in a non-periodic fashion~\cite[e.g. Fig~1 of][]{Costa2011}. \cite{Balsley2004} discussed further details of such episodic development of bursting turbulence, and demonstrated ten-hour time series of measured data indicating patches of eddies with large intervening areas, where all eddies are suppressed on a length scale that is large compared to the size of the largest eddies~\cite[see also][]{Mahrt89,Mahrt99}.

\begin{figure}
  \centering
  \begin{tabular}{cc}
  \includegraphics[height=5.5cm]{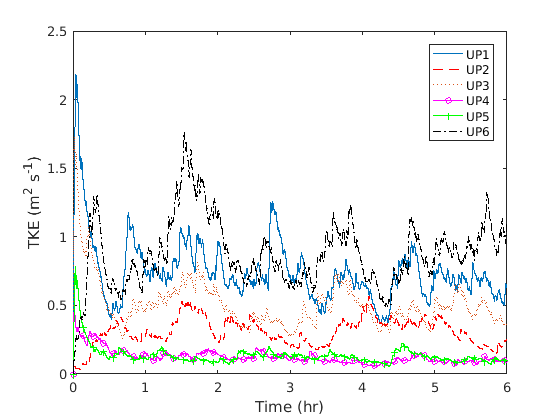}  &
  \includegraphics[height=5.5cm]{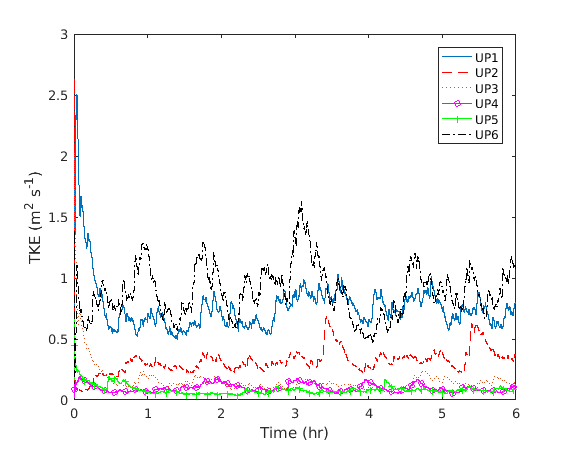}  \\
    $(a)$ Canopy, WALE & $(b)$ CFD, WALE\\
  \includegraphics[height=5.5cm]{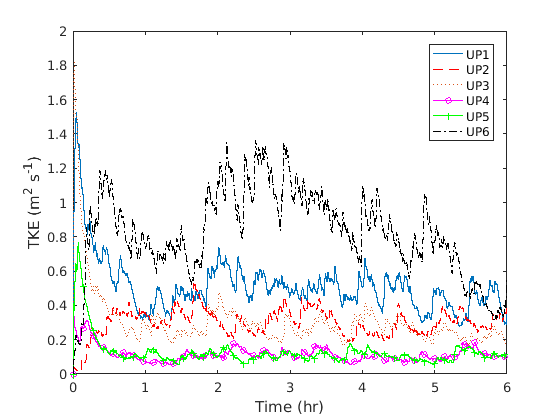}  &
  \includegraphics[height=5.5cm]{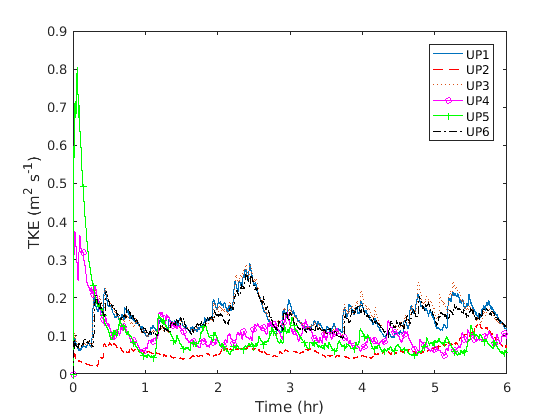}  \\
    $(c)$ Canopy, Smagorinsky & $(d)$ Canopy, WALE, (shallow)\\
  \end{tabular}
  \caption{Time series of turbulence kinetic energy computed at five locations around the obstacle. $(a)$~Canopy model, $(b)$~CFD model.}
  \label{fig:tke}
\end{figure}

\begin{figure}
  \centering
  \begin{tabular}{cc}
  \includegraphics[height=5.5cm]{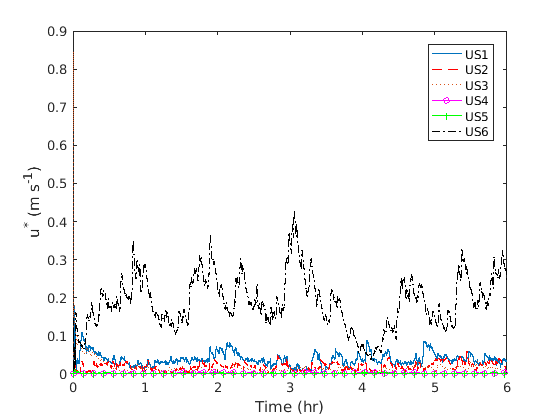}  &
  \includegraphics[height=5.5cm]{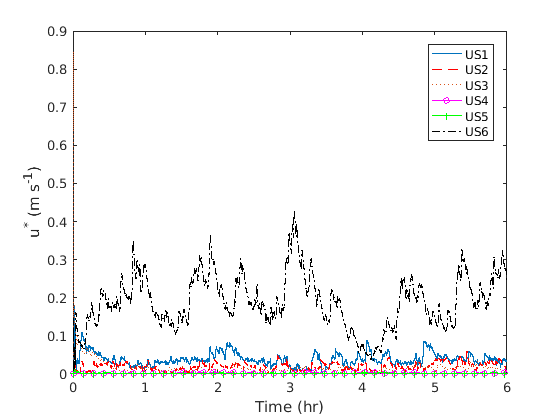}  \\
    $(a)$ LES-WALE, Canopy & $(b)$ LES-WALE, CFD\\
  \includegraphics[height=5.5cm]{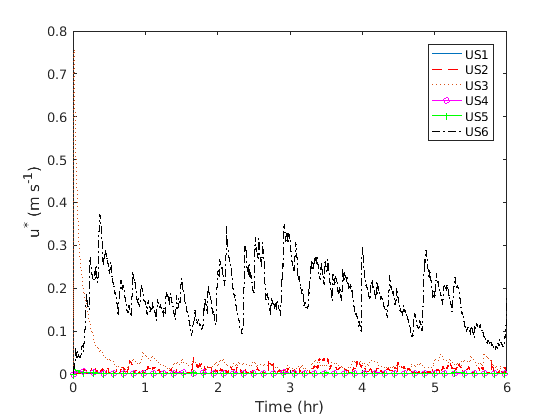}  &
  \includegraphics[height=5.5cm]{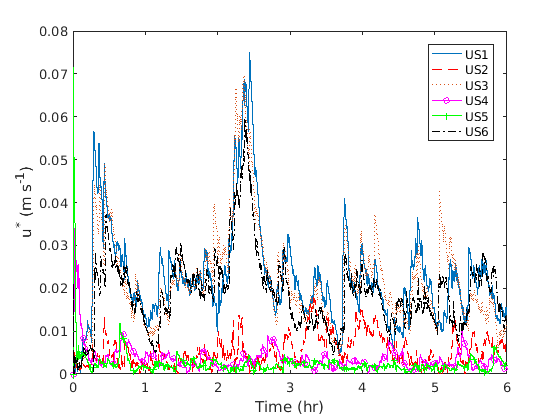}  \\
    $(c)$ LES-Smagorinsky, Canopy & $(d)$ LES-WALE, Canopy, (shallow)\\
  \end{tabular}
  \caption{Time series of the friction velocity experienced by the obstacle. $(a)$~LES-WALE using the canopy method, $(b)$~LES-WALE using the CFD method, $(c)$~LES-Smagorinsky using the canopy method, $(d)$~LES-WALE using the canopy method for the block with $H=75$~m.}
  \label{fig:ustr}
\end{figure}

\begin{figure}
  \centering
  \begin{tabular}{cc}
  \includegraphics[height=5.5cm]{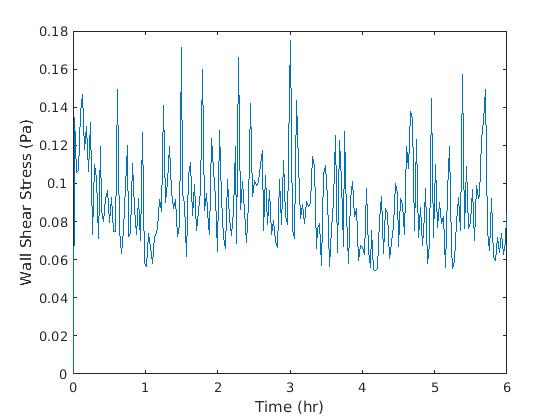}  &
  \includegraphics[height=5.5cm]{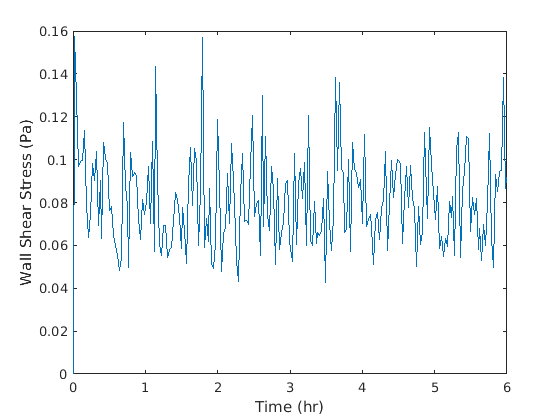}  \\
    $(a)$ LES-WALE, Canopy & $(b)$ LES-WALE, CFD\\
  \includegraphics[height=5.5cm]{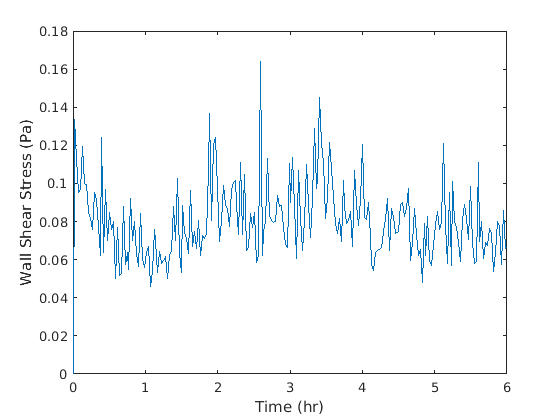}  &
  \includegraphics[height=5.5cm]{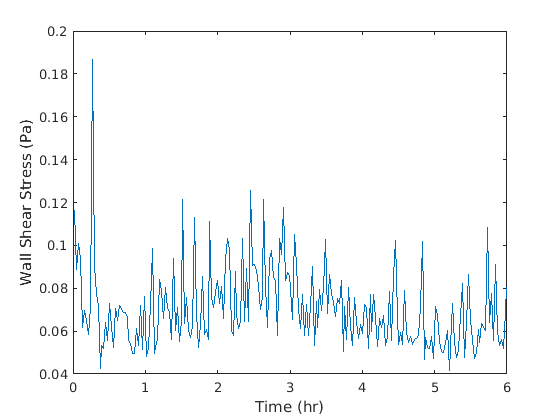}  \\
    $(c)$ LES-Smagorinsky, Canopy & $(d)$ LES-WALE, Canopy, (shallow)\\
  \end{tabular}
  \caption{Time series of the friction velocity $u_{\tau}$ on the ground. $(a)$~LES-WALE using the canopy method, $(b)$~LES-WALE using the CFD method, $(c)$~LES-Smagorinsky using the canopy method, $(d)$~LES-WALE using the canopy method for the block with $H=75$~m.}
  \label{fig:utau}
\end{figure}

To study the bursting events of coherent turbulence in an urban boundary layer under neutral stratification, let us consider the time series of the turbulent kinetic energy. The total kinetic energy $e$ of turbulent eddies may be decomposed into a mean $\bar e$ and a fluctuating part $e'$, where
$$
e = \frac{1}{2}u_iu_i,\quad \bar e = \frac{1}{2}\bar u_i\bar u_i, \quad e' = \frac{1}{2}u'_iu'_i,\quad\hbox{and}\quad\tau'_{ij}=\overline{\rho u'_iu'_j} 
$$
are total energy, mean energy, fluctuating energy, and the Reynolds stress tensor.
The mean of the fluctuating energy over a time window,  {\em i.e.} $\bar e'$  estimates the trace of the Reynolds stress tensor, which is more representative for the overall flow, and given by the equation
$$
\frac{\partial\bar e'}{\partial t} + \bar{u_j}\frac{\partial\bar e'}{\partial x_j} = -\frac{1}{\rho_0}\frac{\partial \overline{u'_ip'}}{\partial x_i} 
-\frac{1}{2}\frac{\partial\overline{u'_ju'_ju'_i}}{\partial x_i} + \nu\frac{\partial^2\bar e'}{\partial x^2_j}
-\overline{u'_iu'_j}\frac{\partial u_i}{\partial x_j} - \nu\overline{\frac{\partial u'_i}{\partial x_j}\frac{\partial u'_j}{\partial x_i}}.
$$
The temporal evolution of the turbulence kinetic energy averaged over a time window and evaluated at six spatial locations was recorded in order to identify the turbulence interaction between the air and the roughness elements. In other words, the strength of a turbulent eddy passing through an assigned location has been recorded instantaneously. The coordinates of these locations are denoted by UP1~(300, 300, 150), UP2~(300, 300, 225), UP3~(380, 300, 150), UP4~(112.5, 300, 0), UP5~(487.5, 300, 0), and UP6~(487.5, 300, 150), where UP1 and UP6 are at the same height as the roughness element with UP6 at a downstream location (see Table~\ref{tab:ups}). The time series calculated by the Canopy Stress model is compared with that of the CFD model. As it is seen from Fig~\ref{fig:tke}, the strength of the eddies at each location features bursting events, and also remains statistically stationary for the entire period of simulation except for an initial spin-up time. %
The spikes of the plots in Fig~\ref{fig:tke} indicate an intermittent transfer of kinetic energy into smaller eddies, a process that leads to episodes of turbulent bursts -- as it is seen from these plots. The time series indicate that a correct amount of momentum has been transferred by the canopy stress model when eddies interacted with the roughness elements. Similarly, the time series of friction velocities $u_*$ and $u_{\tau}$ are presented in Fig~\ref{fig:ustr} and Fig~\ref{fig:utau}, respectively. 


\subsection{Vertical profiles of mean velocity and turbulence stress}
\begin{figure}
  \centering
  \begin{tabular}{cc}
    $(a)$ & $(b)$\\
    \includegraphics[height=5cm]{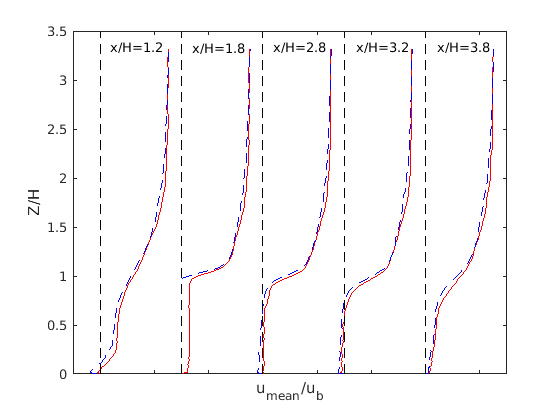}&
    \includegraphics[height=5cm]{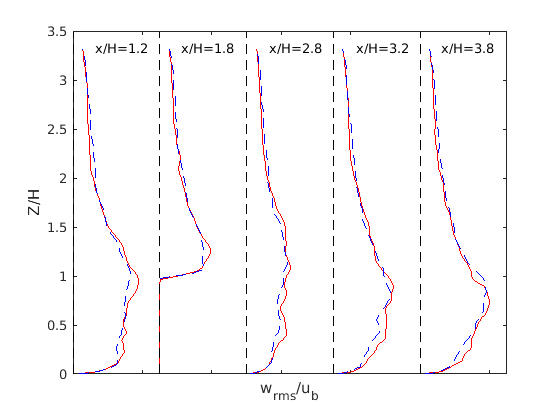}\\
  \end{tabular}
  \caption{The mean stream-wise velocity $\bar u/U_b$ and root mean squared turbulence fluctuations $w_{\hbox{rms}}/U_b$ (wall normal)  as a function of the vertical coordinate at five stream-wise locations: $x/H=1.2$, $x/H=1.8$, $x/H=2.8$, $x/H=3.2$, and $x/H=3.8$. $(a)$~ $\bar u/U_b$; $(b)$ $w_{\hbox{rms}}/U_b$. The results have been compared between the canopy model (solid line) and the CFD model (broken line).}
  \label{fig:vprf}
\end{figure}

\begin{figure}
  \centering
  \begin{tabular}{cc}
    \includegraphics[height=5cm]{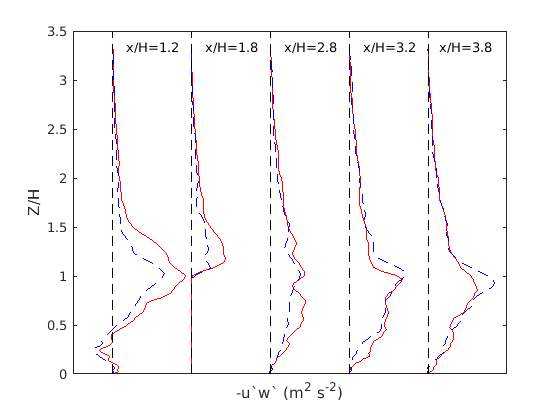}
  \end{tabular}
  \caption{The mean stream-wise velocity $\bar u/U_b$ and root mean squared turbulence fluctuations $w_{\hbox{rms}}/U_b$ (wall normal)  as a function of the vertical coordinate at five stream-wise locations: $x/H=1.2$, $x/H=1.8$, $x/H=2.8$, $x/H=3.2$, and $x/H=3.8$. $(a)$~ $\bar u/U_b$; $(b)$ $w_{\hbox{rms}}/U_b$. The results have been compared between the canopy model (solid line) and the CFD model (broken line).}
  \label{fig:tvprf}
\end{figure}

Vertical profiles of the mean stream-wise velocity and Reynolds stress have been examined and some of the results taken at five stream-wise locations along the center of the domain  have been demonstrated in Fig~\ref{fig:vprf} and Fig~\ref{fig:tvprf}, where the mean and the Reynolds stress are calculated with respect to the average over a fixed time window. LES-WALE prediction of the vertical profiles are compared between the canopy method and the CFD method, where an excellent agreement is observed in Fig~\ref{fig:vprf}$(a)$. The r.m.s. vertical velocity is obtained from the third element on the diagonal of the Reynolds stress tensor and the vertical profiles of which taken at five stream-wise locations are shown in Fig~\ref{fig:vprf}$(b)$. It is evident that both the mean and the turbulent component of the velocity was predicted by LES-canopy model as accurately as it was predicted by the conforming mesh CFD method. In other words, the present extension of the canopy stress method that was also suggested by~\cite{Brown2001}, demonstrates accurate prediction of turbulence for the large-eddy simulation of the flow in an idealized UABL. It is worth mentioning that \cite{Brown2001} found the LES-canopy prediction of turbulence over ridges of varying height demonstrating a much better agreement with the wind tunnel data in comparison with a mixing-length model.

\section{Final discussion and conclusions}\label{sec:con}
LES of urban-contaminant transport typically performs better than operational atmospheric models~\cite[][]{Castro2006,Goodfriend2016}, but pays  a high cost for the numerical resolution because LES resolves a relatively large fraction of the energetic eddies within the surface layer. Strategies to mitigate the computational cost of such an atmospheric LES include the wall modelling approach~\cite[e.g.][]{Senocak2007,Chung2009} that models the effects of energetic scales by modifying the SFS scheme within the surface layer~\cite[e.g.][]{Senocak2007}. One may also consider a locally refined adaptive mesh~\cite[e.g.][]{Castro2006} within the surface layer~\cite[][]{Alam2015,Goodfriend2015,Goodfriend2016} to optimize the cost of atmospheric LES. However, recent work found that LES results were contaminated with errors at the grid refinement interfaces~\cite[][]{Goodfriend2015} if LES is combined with an adaptive mesh. Land-surface parameterization schemes~\cite[][]{Grimmond2011} -- although promising for atmospheric models -- cannot resolve sub-filter scale mixing and transport within an urban canopy~\cite[][]{Cheng2002,Coceal2004}. The present work investigates a canopy stress methodology to simulate the SFS stress exerted by the building-like roughness elements. In this approach, it is not necessary to resolve the complexity of the surfaces of the roughness elements. The stress experienced by air parcels passing over the roughness elements is assumed horizontally homogeneous, and the vertical gradient of this stress is added to the momentum equations. Satisfactory results of turbulent flow over a periodic array of roughness elements indicate that this method is accurate for LES of urban atmospheric boundary layer. 

The accuracy of the canopy stress approach is assessed with respect to a classical CFD approach that implements the no-slip boundary conditions directly on the surface of the roughness elements. Vertical profiles of the mean velocity and its turbulent fluctuations predicted by both methods are in a good agreement. An interesting observation is the richness of the coherent structures which indicates that a large fraction of the unresolved eddies have been captured by the canopy stress method. 

In contrast, \cite{Senocak2007} suggested that a fraction of the unresolved energetic scales within the surface layer can be resolved through the canopy stress method, and \cite{Brown2001} used wind tunnel measurements to argue that the dynamical interaction between the air and the urban roughness elements can be modelled accurately with the canopy stress. In the present work, we explore the canopy stress method proposed by~\cite{Brown2001} to achieve the boundary condition on the surface of an array of building-like obstacles. We have investigated a wall modelling approach in the surface layer that is similar, in principle, to the approach considered by~\cite{Senocak2007}. However, for LES of flow over an array of building-like obstacles, instead of using a canopy stress model within the surface layer, we have considered the wall adaptive local eddy viscosity method~\cite[][]{Nicoud99} to adjust the turbulent stress dynamically in order to resolve a faction of the energetic eddies within the surface layer.

In the present study, a quantitative assessment of the canopy stress method for modelling the dynamics of urban roughness elements has been achieved with respect to a classical CFD technique that implements the no-slip boundary conditions on the roughness elements. The good agreement between the canopy stress model and the CFD model infers the strength of the urban canopy model that avoids the huge cost of capturing unnecessary details of the individual buildings~\cite[][]{Brown2001,Coceal2004}. The Reynolds number $\mathcal Re = 2\times 10^7$ of the present simulation is sufficiently large compared to $3.8\times 10^3$ that was used by~\cite{Goodfriend2016} in a simulation with equivalent geometrical setting. The large-scale intermittency of turbulence exhibited in our simulation supports the conjecture of collapsing and bursting events in atmospheric turbulence~\cite[][]{Mahrt89,Mahrt99}. Until now, large-scale intermittency was attributed to turbulence in the stable (or very stable) atmospheric boundary layer that usually occurs during the night or when the ground is cooled~\cite[][]{Costa2011}. Large-scale intermittency is also an integral feature of atmospheric turbulence at high altitudes under clear sky, which is experienced by aircrafts for about $3\%$ of their cruise time~\cite[][]{Williams2013}, and is more frequent when flying over the relatively urbanized regions~\cite[][]{Jaeger2007}. 
A complete understanding of the large-scale intermittency of atmospheric turbulence in and above urban boundary layers remains elusive, which is important for a number of applications, such as mesoscale weather process, health and comfort of urban dwellers, wind energy technology, and clear air turbulence. Advancing the canopy stress method for LES modelling of atmospheric turbulence in urban boundary layers of real world's complexity is currently underway. 

\section*{Acknowledgements}
The first author acknowledges the Discovery Grant and the second author acknowledges the Undergraduate Student Research Award from the National Science and Research Council~(NSERC), Canada.  The High Performance Computing (HPC) facility utilized for this research work include the Graham cluster of Compute Canada and the CHIA cluster of Center for Health Informatics and Analytics of Memorial University.

\bibliographystyle{model1-num-names}

\end{document}